\documentclass[12pt]{iopart}

\usepackage[dvips]{graphicx}
\usepackage{iopams}
\usepackage{hyperref}
\begin{document}

\sloppy

\title[A new model for simulating colloidal dynamics]
{A new model for simulating colloidal dynamics}

\author{Vladimir Lobaskin\footnote[1]{
To whom correspondence should be addressed 
(lobaskin@mpip-mainz.mpg.de)} 
and Burkhard D\"unweg}

\address{Max Planck Institute for Polymer Research, 
Ackermannweg 10,
D-55128 Mainz, Germany}

\begin{abstract}
We present a new hybrid lattice-Boltzmann and Langevin molecular
dynamics scheme for simulating the dynamics of suspensions of
spherical colloidal particles. The solvent is modeled on the level of
the lattice-Boltzmann method while the molecular dynamics is done for
the solute. The coupling between the two is implemented through a
frictional force acting both on the solvent and on the solute, which
depends on the relative velocity. A spherical colloidal particle is
represented by interaction sites at its surface. We demonstrate that
this scheme quantitatively reproduces the translational and rotational
diffusion of a neutral spherical particle in a liquid and show
preliminary results for a charged spherical particle. We argue that
this method is especially advantageous in the case of charged
colloids.
\end{abstract}


\submitto{\NJP}

\maketitle

\section{Introduction}
\label{sec:intro}

Understanding the dynamics of colloidal dispersions has been a
long-standing problem in condensed matter physics. Despite the
continuous need for accurate theoretical predictions concerning
particle mobilities or sedimentation rates, the development of the
field meets a nearly unpassable barrier. The many-body character of
the hydrodynamic interactions (HI) between the colloidal particles
(i.~e. their correlated motion as a result of solvent-mediated fast
momentum transport) poses a serious challenge for analytic theory. On
the other hand, the rapid growth of available computer power in the
last decades has made it more and more feasible to descend to a more
basic description of colloidal systems rather than trying to develop
an adequate macroscopic description. Such a step was made recently 
for the statics of charged colloids, where simulations of the
primitive electrolyte model, taking the particle character of the
charges and fluctuations in the charge distribution fully into
acount, resolved some important strong correlation issues and thus
have resulted in a significant refinement of the existing theories of
screening \cite{Hansen1}. In a similar fashion, many-body effects in
the HI of suspensions of only moderate density are quite
important. Although their analytical form is, in principle, known in
terms of a multipole expansion \cite{Mazur}, it is numerically very
cumbersome to take these higher-order many-body terms into account in
a Brownian or Stokesian dynamics \cite{Brady} simulation, where only
the positions of the colloidal particles enter. For this reason, it is
more convenient and for large particle numbers also more efficient,
to take these effects into account by simulating the solvent degrees
of freedom, and, in particular, the momentum transport through the solvent
explicitly. An additional bonus is that retardation effects, if
present, are automatically taken into account.

Standard Molecular Dynamics (MD), where just Newton's equations of
motion are integrated numerically, is of course able to simulate HI
correctly. However, it is very time-consuming to simulate the motion
of a huge number of particles explicitly in such detail. The time step
is governed by the local oscillations of the solvent particles in
their temporary ``cages'', which is much faster than the motion of the
colloidal particles. This information is however not needed for
correctly reproducing HI. One just needs an appropriate mechanism for
momentum transfer on somewhat larger time scales (however still small
compared to the motion of the colloidal particles). This can be done
in various ways: Dissipative Particle Dynamics (DPD) \cite{Warren} is
essentially MD, where however the particles are made very soft in
order to increase the time step, and where a momentum-conserving
Langevin thermostat is added \cite{Thoso}. Another possibility is multi-particle 
collision dynamics \cite{Malevanets}, where the solvent is modeled as
an ideal gas, and collisions are modeled as simple stochastic updating
rules which conserve energy and momentum. Grid-based methods can be
either the direct solution of the Navier-Stokes equation by a finite
difference method, or the Lattice Boltzmann (LB) \cite{Succi}
approach, where essentially a linearized Boltzmann equation is solved
in a fully discretized version (i.~e. space, time, and velocities are
discretized).

In our opinion, all of these approaches have their advantages and
disadvantages. Nevertheless, it seems that all of them are roughly
equivalent with respect to their computational efficiency, and with
respect to the level of accuracy of describing the physical
phenomena. Therefore, we believe that the choice of method is mainly a
matter of convenience.

For our model we have chosen the LB route, mainly because this
algorithm is particularly simple to implement and parallelize. Moreover, we were inspired by previous highly successful simulations of hard-sphere colloidal suspensions \cite{Ladd0,Ladd1,Ladd2,Ladd3,Ladd4,Frenkel1,Frenkel2,Frenkel3,Frenkel4}. A further advantage of a grid-based method is that thermal fluctuations can be both turned on and off, depending on the physical situation under consideration (in a particle method they are always present). This approach is essentially a hybrid method, where the solvent is modeled via LB, while the colloidal particles are run with MD.

The original approach by Ladd \cite{Ladd0,Ladd1,Ladd2,Ladd3,Ladd4}
models the colloidal particles as extended hollow spheres, while stick
boundary conditions at the surface are implemented (roughly spoken)
via bounce-back collision rules. One should note that for moving spheres
this involves some minor fluctuations 
in the fluid mass, which is included within a sphere: The moving
shell incorporates some new fluid at the front, while it releases
mass at the rear. The fluctuations are a natural effect of the
thermal density fluctuations of the fluid, and of the lattice
discretization. Using this method, the translational and the rotational
dynamics of colloidal spheres were accurately described.

This model has been recently extended to the case of charged systems
\cite{Horbach1}, where the colloidal particles carry a charge, while
the counterions and salt ions are taken into account as LB
populations, such that the electrostatics is essentially taken into
account on the level of the Poisson-Boltzmann equation. This method
has two disadvantages: firstly, the discrete nature of the ions and
correlations beyond the Poisson-Boltzmann level are not taken into
account; secondly, one cannot avoid a leakage of charge into (and out 
of) the sphere (as in the case of mass), such that it is hard
(if not impossible) to maintain a well-defined Debye layer of charges
around it \cite{Horbach2}. For these reasons, it is advisable
to take the counterions explicitly into account. The disadvantage,
however, is that only rather small size ratios (size of colloidal
particle vs. size of ion) are easily accessible. Nevertheless, we
believe it is useful to study such a system, with respect to both its
equilibrium and its nonequilibrium properties. The purpose of the
present paper is to describe first steps in the development and
validation of the corresponding model.

For the coupling of the small ions to the LB hydrodynamics, one would
like to simply model the former as point particles. Fortunately, such
a coupling has been recently developed by Ahlrichs and D\"unweg
\cite{Ahlrichs1,Ahlrichs2} with the purpose of studying the dynamics
of polymer solutions \cite{Ahlrichs3}. Each Brownian point particle is
assigned a phenomenological friction coefficient $\zeta$, and the
coupling is implemented just as a dissipative force $\vec F = - \zeta
(\vec V - \vec u)$ acting on the particle, where $\vec V$ is the
particle velocity, while $\vec u$ is the solvent flow velocity at the
particle's position, obtained via linear interpolation from the
surrounding lattice sites. Furthermore, the LB variables at these
sites are adjusted to ensure momentum balance, and thermal
fluctuations are added to both types of degrees of freedom. For more
details on implementation and validation of this method, see
Refs. \cite{Ahlrichs1,Ahlrichs2}.

It is then convenient to use this coupling not only for the ions, but
also for the colloidal particles, whose larger size is taken into
account by modeling them as some arrangement of interaction sites. For
reasons of efficiency, we take only points at the surface of the
colloid. Furthermore, the coupling constant $\zeta$ is chosen rather
large, which approximates stick boundary conditions. It should be
noted that the dissipative nature of the coupling does not model any
squeezing-out of solvent, hence the sphere is filled with the same
amount of solvent regardless if it is hollow or if it were filled with
particles. The solvent within the sphere follows the motion of the
surrounding shell (with some minor time lag, see below), and therefore
one can view the solvent inside the sphere as just belonging to the
colloidal particle. Adding further particles in the sphere's volume
would have no effect except coupling the fluid within the sphere even
tighter to its motion. Within our desired level of accuracy, this
turned out not to be necessary. The long-time motion of the sphere is
the same as that of a corresponding hard sphere, as we will show in
the present paper.

We analyze basic dynamic properties of such a model colloid, where we
restrict ourselves to the case of a single sphere. A detailed analysis
is done for the neutral case, while some preliminary data for the case
of a charged sphere are presented. We show that the (neutral) model
exhibits the essential dynamical features of a spherical colloidal
particle in a liquid. In our view, our method provides comparable
efficiency to Ladd's approach, while being quite straightforward to
implement. Furthermore, it provides substantial flexibility with
respect to the properties of the colloidal surface, namely, deformable,
permeable, and non-stick surfaces can be easily simulated.

The remainder of this article is organized as follows: In
Sec. \ref{sec:model}, we describe our simulation model, while
Sec. \ref{sec:results} contains the numerical results on translational
and rotational diffusion. Finally, Sec. \ref{sec:summary} concludes
with a brief summary.

\section{Model}
\label{sec:model}

Our hybrid simulation method involves two subsystems: the solvent that
is modeled via LB with fluctuating stress tensor (i.~e. we run a
constant-temperature version of the LB method) and a Langevin MD
simulation for the particles immersed in the solvent. The LB
simulation is performed using the 18-velocity model \cite{Ladd1},
using the protocol described in \cite{Ahlrichs1,Ahlrichs2}. The fluid
simulation consists of collision and propagation steps, the former
being performed with inclusion of the momentum transfer from the
solute particles (surface beads, and, for charged systems, ions).

\begin{figure}
\begin{center}
\vskip 0.1in
\includegraphics[clip,width=0.5\textwidth]{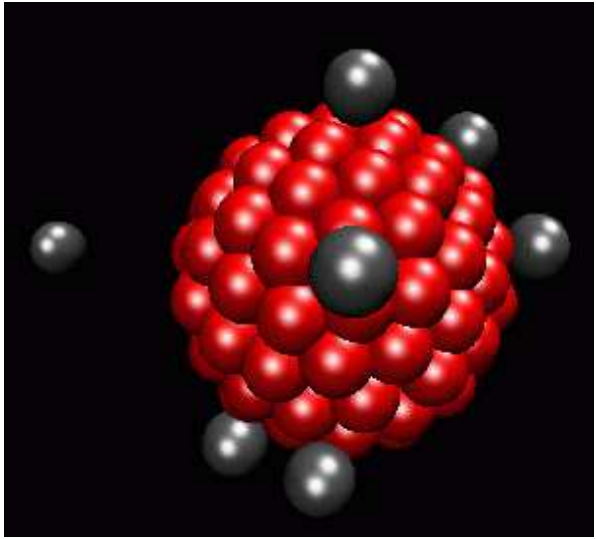}
\end{center}
\caption{
Raspberry-like model of a colloidal sphere. There is a central large
bead of radius $R=3$ and charge $Z=10$. The small beads of radius 1
are connected with their nearest neighbors on the surface via FENE
bonds. A repulsive soft-core potential is also operating between all
the monomers. The counterions are moving freely in space and interact
with the central bead via the Coulomb potential and with the surface
beads via the repulsive LJ potential. 
\href{http://www.mpip-mainz.mpg.de/~lobaskin/TR6/Z20.mpg}{Movie}
}
\label{fig:sphere}
\end{figure}

The colloidal particle is represented by a two-dimensional tethered
bead-spring network consisting of $100$ beads, which is wrapped around
a ball of a radius $\sigma_{cs}$ (for notation, see below), so that
the whole construction resembles a raspberry (see
Fig. \ref{fig:sphere}). The network connectivity is maintained via
finitely extendable nonlinear elastic (FENE) springs,
\begin{equation}
V_{FENE} (r) = - \frac{k R_0^2}{2} \ln
\left(1 - \left( \frac{r}{R_0}\right)^{2} \right) ,
\label{eq:fene}
\end{equation}
where $k$ is the spring constant, and $R_0$ the maximum bond
extension. Furthermore, the beads repel each other by a modified
Lennard-Jones (LJ) potential
\begin{equation}
V_{LJ} (r) = \left\lbrace \begin{array} {ll} 
4\epsilon_{ij} \left(  \left( \frac{\sigma_{ij}}{r} \right)^{12} 
- \left( \frac{\sigma_{ij}}{r} \right)^{6} + 
\frac{1}{4} \right) & r<2^{1/6} \sigma_{ij}  \\
0  & r\ge 2^{1/6} \sigma_{ij} . \end{array} \right.
\label{eq:lj}
\end{equation}
An additional repulsive LJ bead is introduced at the center of the
sphere in order to maintain its shape. In Eq. \ref{eq:lj}, $i,j$
denote either a central (``c'') or a surface (``s'') bead. The unit
system is completely defined by the surface bead parameters by setting
$\epsilon_{ss}$, $\sigma_{ss}$, and the surface bead mass $m_s$ to
unity. The interaction between the central bead and the surface beads
is described by $\sigma_{cs} = 3$, which is thus the sphere radius,
and $\epsilon_{cs}=8$. Furthermore, the FENE spring constant for the
surface beads is $k=300$ and the maximum bond extension is $R_0=
1.25$. To simulate a charged colloidal particle, we place the charge
at the central bead, and add an appropriate number of counterions (LJ
beads with ``s'' properties) outside the sphere. The electrostatic
interaction is taken into account via the Coulomb potential
\begin{equation}
V_{el} (r) = \lambda_B k_B T \frac{q_i q_j}{r}
\label{eq:electrostatics}
\end{equation}
between the various charges, where the standard Ewald summation
technique \cite{Allen} is applied. In Eq. \ref{eq:electrostatics},
$\lambda_B=e^2/ \left(4\pi\varepsilon_0 \varepsilon_r k_B T\right)$ is
the Bjerrum length, $k_B$ the Boltzmann constant, $q_i$ the charge of
species $i$ in units of the elementary charge $e$, and $T$ the
temperature.

The LB lattice constant is chosen as one (in our LJ unit system), and
the fluid is simulated in a cubic box with periodic boundary
conditions. The force between the LB fluid and the surface beads
(or ions) is given by
\begin{equation}
\vec{F}= - \zeta \left( \vec{V} - \vec{u} \right) + \vec{f} .
\label{eq:4}
\end{equation}
Here, $\zeta$ is the ``bare'' \cite{Ahlrichs2} friction coefficient,
$\vec{V}$ and $\vec{u}$ are the velocities of the bead and the fluid
(at the position of the bead), respectively, while $\vec{f}$ is a
Gaussian white noise force with zero mean, whose strength is given via
the standard fluctuation-dissipation theorem
\cite{Ahlrichs1,Ahlrichs2} to keep the surface beads and ions at
the same temperature as the solvent. The central bead is not
coupled to the solvent (here $\zeta = \vec f = 0$); as discussed in the
Introduction, the behavior of the model would change only marginally
if such a coupling were included. In our simulation we used a friction
constant $\zeta = 20$, a temperature $k_B T = 1$, a fluid mass density
$\rho = 0.85$, and a kinematic viscosity $\nu=3$, resulting in a
dynamic viscosity $\eta=2.55$. At least 20000 MD steps were performed
to equilibrate the initial random bead configuration before the
interaction with the LB solvent was turned on. Further details on the
method can be found in \cite{Ahlrichs1,Ahlrichs2}.

\section{Results}
\label{sec:results}

\begin{figure}
\vskip 0.1in
\begin{center}
\includegraphics[clip, width=0.75\textwidth]{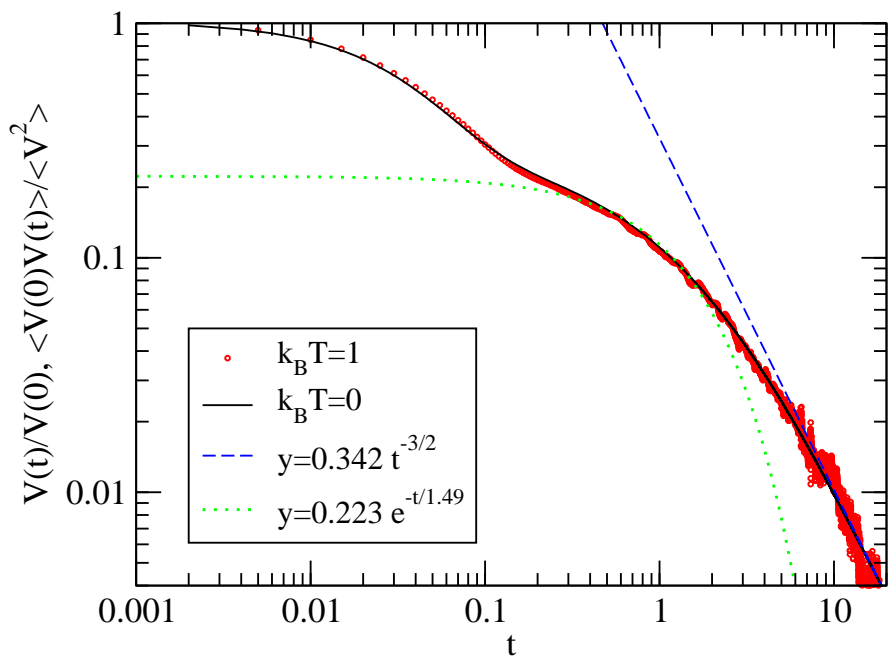}
\end{center}
\caption{
Normalized translational velocity of the center of mass of the
colloidal sphere in a "kick" experiment at $k_B T = 0$, and its
normalized velocity autocorrelation function for Brownian motion at
$k_B T = 1$. The dotted curve shows the exponential decay derived from
the Stokes law and the dashed curve the expected long-time asymptotic
behavior.}
\label{fig:vcf2}
\end{figure}

We test the simulation method against basic relations for an isolated
sphere in solvent. First, we look at the center of mass' velocity
relaxation. The simplest experiment to perform is a ``kick''. The
sphere is placed in a LB fluid at rest (i.~e. without thermal noise),
and at time $t = 0$ all particles of the sphere are assigned an
identical velocity $\vec V = 1$ in $x$ direction. Fig. \ref{fig:vcf2}
monitors the time behavior of the sphere's center of mass velocity,
normalized by the initial value. According to linear response theory,
this relaxation function must be identical to the normalized
center-of-mass velocity autocorrelation function for Brownian motion
in thermal equilibrium, if the initial kick is weak enough. This is
indeed satisfied, as a comparison of the two curves in
Fig. \ref{fig:vcf2} shows. For the experiment in thermal equilibrium,
we performed 10 runs with different random number generator
initializations, in order to reduce the statistical uncertainty.

It is well-known that simulations of Brownian motion in a hydrodynamic
solvent are always strongly affected by finite size effects. The
diffusion constant, and therefore also the relaxation function, depend
on the linear system size $L$ due to hydrodynamic interactions with
the periodic images. The diffusion constant exhibits a finite-size
correction of order $R / L$ \cite{Ahlrichs2,Burkhard}. Asymptotic
behavior can therefore only be expected for $R / L \ll 1$, and this is
why we performed the experiment in a rather large box of size $L =
80$. For the same reason, the equivalence between ``kick'' experiment
and Brownian motion will only hold if the comparison is done for the
same box sizes.

\begin{figure}
\begin{center}
\vskip 0.1in
\includegraphics[clip, width=0.75\textwidth]{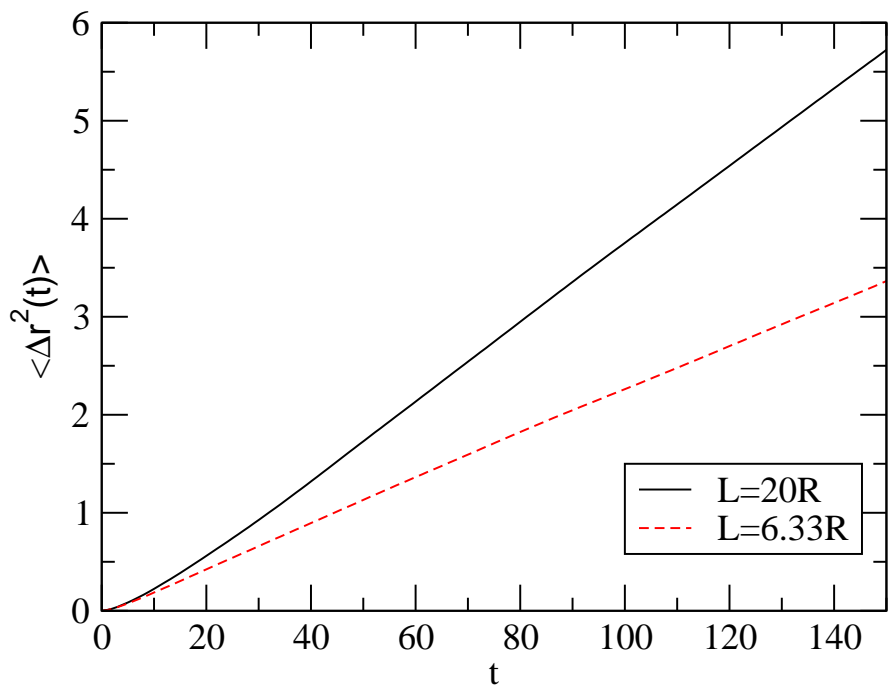}
\end{center}
\caption{
Mean-square displacement of the center of mass of a neutral sphere at
$k_B T = 1$ and different simulation box sizes. }
\label{fig:msd}
\end{figure}

One clearly sees that the initial decay of the relaxation function is
characterized by two relaxation processes, one initial fast decay
followed by a somewhat slower relaxation. Qualitatively, this may be
explained as follows: A compact sphere of radius $R$ and mass $M$
should exhibit a velocity relaxation which, after transient ballistic
motion, is initially characterized by an exponential decay $\exp
\left( - t / \tau \right)$, with a relaxation time $\tau = M /
\zeta_{tot}$. Here, $\zeta_{tot}$ is the total friction
coefficient, which we estimate via Stokes' law for stick boundary
conditions as $\zeta_{tot} \approx 6 \pi \eta R \approx 144$. However,
the effective mass is time-dependent: While initially only the mass of
the beads $M = 101$ contributes, at later times the fluid within the
sphere is dragged as well, such that then the mass is roughly
estimated as $M \approx 101 + 4 \pi \rho R^3 / 3 \approx 214$.  This
gives rise to the initial and final relaxation times $\tau_{in}
\approx 0.7$ and $\tau_{fin} \approx 1.5$. However, the initial
relaxation time cannot be observed, since in the extreme short-time
regime ballistic effects play a role. Conversely, the decay with
$\tau_{fin}$ is clearly visible (see Fig. \ref{fig:vcf2}).

\begin{figure}
\begin{center}
\vskip 0.1in
\includegraphics[clip, width=0.75\textwidth]{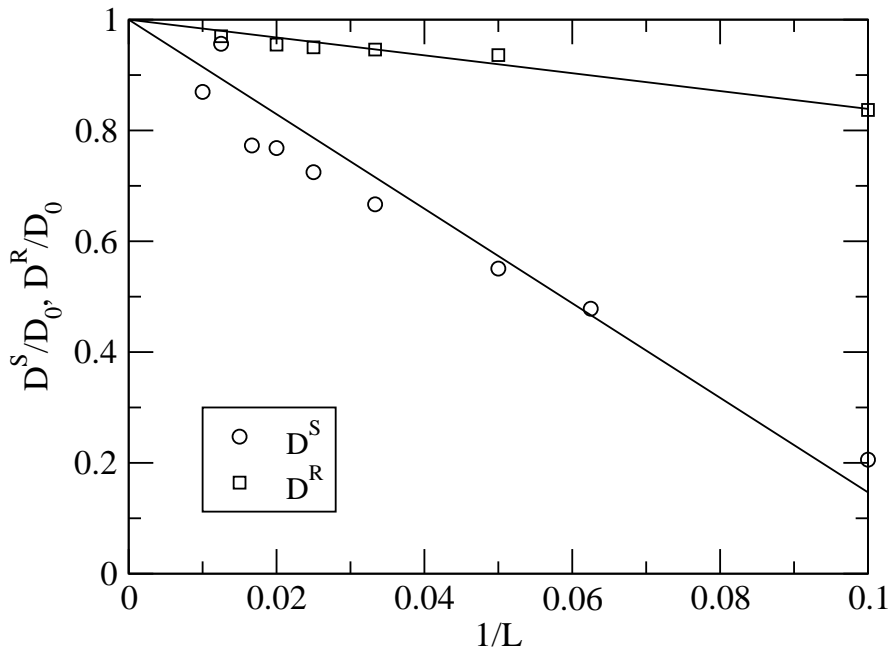}
\end{center}
\caption{
The long-time self-diffusion coefficient and the rotational diffusion
coefficient of the colloidal sphere (normalized by the asymptotic
Stokes-Einstein and Stokes-Einstein-Debye values) as functions of the inverse size of the primary simulation cell. }
\label{fig:ds}
\end{figure}

After $t \approx 1$, the famous long-time tail \cite{Alder,Hansen2}
(normalized relaxation function $V(t) / V(0) = B t^{-3/2}$) sets
in. The physical mechanism of this slow relaxation is the fact that
the momentum is conserved and hence transported away diffusively from
the particle \cite{Hansen2}. For this reason, it can only be observed
in a sufficiently large box. The prefactor of the power law is known
in the colloidal limit, $B = (1 / 12) (N m / \rho) (\pi \nu)^{-3/2}$,
where $N$ is the number of beads, and $m$ the bead mass
\cite{Hinch,Cichocki}. It should be noted that the prefactor of the
power-law decay of the {\em unnormalized} velocity autocorrelation
function does not depend on the properties of the sphere at all, but
only on the temperature, and the hydrodynamic properties of the
solvent \cite{Hinch,Cichocki}. The mass dependence of $B$ results only
from the normalization, i.~e. the value of $\left< V^2 \right>$
according to the equipartition theorem, $\left< V^2 \right> = 3 k_B T
/ (N m)$. From this consideration it is clear that the {\em
short-time} value of the sphere's mass ($N m$) enters (at $t = 0$
the shell and the inner fluid are not yet coupled). For our
simulation parameters, we find $B = 0.342$. As Fig. \ref{fig:vcf2}
shows, the data exhibit the expected behavior very nicely.

Figure \ref{fig:msd} illustrates the finite size effect in the diffusive
properties by plotting the mean square displacement of the colloidal
sphere as a function of time, for two different box sizes $L = 6.33 R$
and $L = 20 R$, for Brownian motion at $k_B T = 1$. In the long-time
regime, the slope is given by $6 D^S$, where $D^S$ is the
self-diffusion coefficient. The figure clearly shows that $D^S$
increases with box size. This can be explained in terms of
hydrodynamic interactions with the periodic images, or, equivalently,
in terms of suppression of long-wavelength hydrodynamic modes
\cite{Ahlrichs2,Burkhard}. We have therefore performed a systematic
finite-size analysis, and measured $D^S$ for various box sizes by
integrating the velocity autocorrelation function over time
\cite{Hansen2}. In Fig. \ref{fig:ds} we plot $D^S / D_0$ 
($D_0$ denoting the asymptotic Stokes-Einstein value $D_0 = k_B T / (6
\pi \eta R) = 6.9 \times 10^{-3}$) as a function of $1 / L$; the
expected linear behavior \cite{Burkhard} is clearly seen.

\begin{figure}
\begin{center}
\vskip 0.1in
\includegraphics[clip, width=0.75\textwidth]{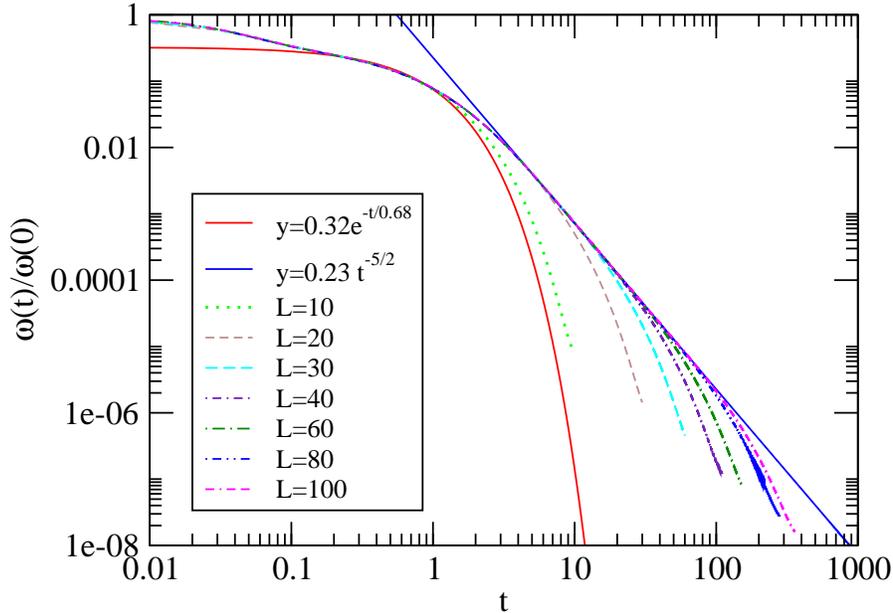}
\end{center}
\caption{
Decay of angular velocity of the colloidal sphere. The different
curves are marked by the primary simulation box sizes. The solid
curves show the expected exponential decay according to the Debye law
and the long-time asymptotic behavior.}
\label{fig:avcf}
\end{figure}

We now look at the relaxation of the rotational motion. We performed a
similar kick experiment as described above, where now an initial
angular velocity $\omega_0 = 1$ was provided to the sphere. The data
for the normalized decay function $\omega(t) / \omega (0)$ are
presented in Fig. \ref{fig:avcf}. The sphere dynamics shows a
characteristic ``raw egg'' (damped) rotation pattern with a fast
initial decay and a subsequent slower one. For the rotational
relaxation of a hard sphere with moment of inertia $I$ and rotational
friction coefficient $8 \pi \eta R^3$, theory predicts a decay
according to the Debye law $\omega (t) / \omega (0) =
\exp \left( - t / \tau_r \right)$, where the relaxation time is given
by $\tau_r = I / (8 \pi \eta R^3)$ \cite{Debye}. Similarly to the
effective mass, the effective moment of inertia is expected to
increase as a function of time, due to the time-delayed dragging of
the fluid. Initially we expect a hollow-sphere value of roughly
$I_{in} \approx (2/3) M R^2 \approx 600$, where we used the mass of
the outer shell $M = 100$. A more accurate value is obtained by
direct summation over the contributing beads, yielding $I_{in} = 546$.
In the late-time regime the moment of
inertia is expected to be the compact-sphere value $I_{fin} \approx
(2/5) M R^2 \approx 770$, where the total mass $M \approx 214$ has
been used. This results in relaxation times $\tau_{r,in} = 0.34$ and
$\tau_{r,fin} = 0.44$. These values are roughly consistent with a fit
to the data in the interval $0.1 < t < 1.0$, which yields a somewhat
larger relaxation time $\tau_r = 0.68$ (see Fig. \ref{fig:avcf}).
Given the general inaccuracy of the
estimates, and the crossover of the Debye relaxation function both to
short-time ballistic behavior, and long-time hydrodynamic behavior,
this deviation is not too surprising.

Similar to the translational motion, the exponential decay is then
followed by a power-law long-time tail. Theory predicts $\omega (t) /
\omega (0) = (\pi I / \rho) (4 \pi \nu t)^{-5/2}$ \cite{Hauge}. As we
have seen before, the long-time tail in the translational case is
governed by the short-time mass as a result of normalization.
Similarly, the rotational tail must be controlled by the {\em
short-time} moment of inertia $I_{in}$, which also governs the mean
square fluctuations of the angular velocity via the equipartition
theorem $\left< \omega^2 \right> = 3 k_B T / I_{in}$. For this reason,
we can determine $I_{in}$ by fitting a $t^{-5/2}$ law to the data,
which is much more accurate than our rough geometrical estimate. The
fit results in $I_{in} = 533$, which is reasonably consistent.  If we
insert this value into the relaxation time expression, we obtain
$\tau_r = 0.64$, in quite good agreement with the data. It hence seems
that fluid dragging effects are not yet very important for the initial
Debye relaxation.

At longer times, one can again notice a significant finite-size
effect: the curves obtained in the smaller simulation box depart from
the asymptotic power-law line earlier, i.~e. the long-time diffusion
is hindered at the small system sizes. In fact, the power-law regime
is inaccessible at $L = 10$ while for $L = 100$ it extends up to $t =
200$. This value characterizes the interval after which the particle
starts feeling its own periodic images. The rotational diffusion
constant $D^R$ is given by the Green-Kubo integral \cite{Berne1}
\begin{equation}
D^R = \frac{1}{3} \int_0^\infty dt\, \left< \vec \omega (t)
\cdot \vec \omega (0) \right> .
\end{equation}
We can evaluate this by again making use of linear response
theory: The correlation function $\left< \vec \omega (t)
\cdot \vec \omega (0) \right>$ is identical to the relaxation
function presented in Fig. \ref{fig:avcf}, multiplied with the initial
value $\left< \omega^2 \right>$, which we know from the equipartition
theorem (see above). Using this approach, we have calculated $D^R$ for
different box sizes $L$. In an infinite fluid, $D^R$ has the
Stokes-Einstein-Debye value $D^R = k_B T / (8 \pi \eta R^3) = 0.58 \times
10^{-3}$, towards which the results indeed converge for $L \to
\infty$. Again a $1 / L$ behavior (but weaker than for translational
diffusion) is observed, as shown in Fig. \ref{fig:ds}. It should be
noted that the accuracy of the data is slightly hampered by the fact
that we had to use a somewhat arbitrary criterion for cutting off the
integral function, which does not arrive at a constant value but
rather ends up with a linear increase originating from a small nonzero
constant angular velocity in the long-time limit (as a manifestation
of the finite-size effect).

\begin{figure}
\begin{center}
\vskip 0.1in
\includegraphics[clip, width=0.75\textwidth]{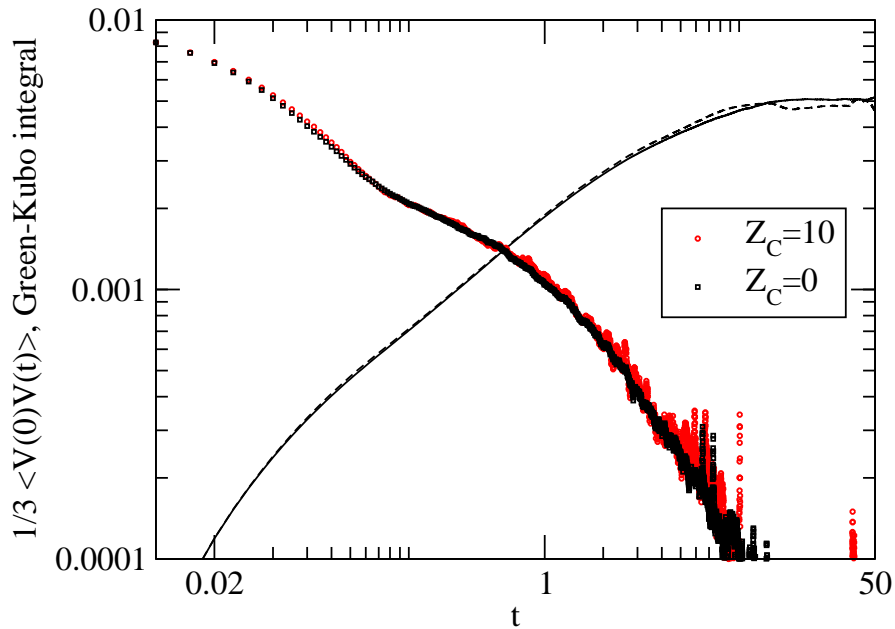}
\end{center}
\caption{
Center of mass velocity autocorrelation functions for a neutral and
a charged colloidal sphere. The Green-Kubo integral function
is also shown as solid curve for $Z=0$ and dashed curve for $Z=10$.
}

\label{fig:zz}
\end{figure}

Finally, we have started to study the effect of charge on the
self-diffusion of the colloidal particle. We performed simulations of
a sphere with central charge $Z_C = 10$ in an LB box of size $L =
40$. Ten counterions of charge $-1$ were also added. The Bjerrum
length was set to $2$. Technically, the simulation ran without
any problems just as well as for the neutral system. In Fig.
\ref{fig:zz} we compare the decay of the velocity
autocorrelation function for the neutral and charged spheres. The
difference between the two is not detectable within our error bars at
this charge. We cannot say yet much about the influence of the charges
on the dynamic properties; we do however expect that for strongly
charged systems the Debye layer will effectively increase the
hydrodynamic radius and slow down the diffusion. More work needs to be
done to resolve this issue.

\section{Summary}
\label{sec:summary}

In summary, we introduced and tested a new model for simulating
colloidal dynamics with inclusion of the hydrodynamic effects on the
level of the lattice-Boltzmann equation. The suggested
``raspberry''-like colloidal object exhibits the essential features of
the diffusion of a spherical colloidal particle. For the first time we
combined a model containing explicit charges with accurate treatment
of the hydrodynamics. We expect this method to have some advantages
over the previously applied scheme \cite{Horbach1} which builds upon
the model of Ladd \cite{Ladd0} and has the problem of charge
leakage. In our method: (i) the stick boundary conditions are realized
not strictly, i.~e. the particle surface is softly bound to the fluid;
(ii) the method is easily adaptable to the multiple timestep method
where the ionic action on the colloid can be averaged out somewhat
within the integration step for the solvent. We should note that such
a hybrid simulation requires still some computational effort --- two
million hybrid steps for the colloid consisting of 100 beads in a box of
$20 \times 20 \times 20$ LB cells take about 12 hours on a 2 GHz Pentium
4 processor, while it is expected to run a few times faster on more
sophisticated hardware. Here, one step is a MD step of the total system
of beads, plus one LB update of the whole lattice, where the latter part
completely dominates the CPU effort. Some improvement is possible by not updating the LB fluid at every MD step, as it was done here. For dense
systems, where the MD part is no longer negligible, one should also
optimize the interaction potentials. Finally, we would like to mention
that the algorithm can be fully parallelized and we are working on the
parallel version.

\ack 
We thank Christian Holm, J\"urgen Horbach and Kurt Kremer
for stimulating discussions, and the latter also for critical
reading of the manuscript. This work was funded by the SFB
TR 6 of the Deutsche Forschungsgemeinschaft.

\end{document}